\documentclass[10pt]{article}

\usepackage{amsmath}
\usepackage{amssymb}

\usepackage{graphicx}

\usepackage{cite}

\usepackage{color} 

\usepackage[labelfont=bf,labelsep=period,justification=raggedright]{caption}

\makeatletter
\renewcommand{\@biblabel}[1]{\quad#1.}
\makeatother

\date{}

\begin{document}

\begin{flushleft}
{\Large
\textbf{Optimal control of transitions between nonequilibrium steady states}
}
\\
Patrick R. Zulkowski$^{1\ast}$, 
David A. Sivak$^{2}$, 
Michael R. DeWeese$^{3}$
\\
\bf{1} Department of Physics, University of California, Berkeley, CA 94720, USA \\
	  Redwood Center for Theoretical Neuroscience, University of California, Berkeley CA 94720, USA
\\
\bf{2} Center for Systems and Synthetic Biology, University of California, San Francisco, CA 94158, USA
\\
\bf{3} Department of Physics, University of California, Berkeley, CA 94720, USA \\
          Redwood Center for Theoretical Neuroscience, University of California, Berkeley CA 94720, USA \\
	  Helen Wills Neuroscience Institute, University of California, Berkeley CA 94720, USA
\\
$\ast$ E-mail: pzulkowski@berkeley.edu
\end{flushleft}

\section*{Abstract}
Biological systems fundamentally exist out of equilibrium in order to preserve organized structures and processes. Many changing cellular conditions can be represented as transitions between nonequilibrium steady states, and organisms have an interest in optimizing such transitions. Using the Hatano-Sasa Y-value, we extend a recently developed geometrical framework for determining optimal protocols so that it can be applied to systems driven from nonequilibrium steady states. We calculate and numerically verify optimal protocols for a colloidal particle dragged through solution by a translating optical trap with two controllable parameters. We offer experimental predictions, specifically that optimal protocols are significantly less costly than naive ones. Optimal protocols similar to these may ultimately point to design principles for biological energy transduction systems and guide the design of artificial molecular machines.


\section*{Introduction}
\label{Introduction}
Living systems are distinguished by their self-organization. Given the entropic driving force embodied in the second law of thermodynamics, creating and maintaining such organization requires staying far from equilibrium~\cite{Schrodinger:WhatIsLife}, typically by coupling to nonequilibrium gradients. For example, ATP-driven molecular motors (\emph{e.g.}, kinesin) are forced away from equilibrium by cellular maintenance of a chemical potential difference between ATP and ADP~\cite{Howard}, and the rotary Fo-F1 ATP synthase operates out of equilibrium due to cellular maintenance of an electrochemical gradient across the inner mitochondrial membrane~\cite{MBOC}.
For constant ATP and ADP concentrations, or constant membrane potential, the dynamics of an ensemble of such molecular motors will approximate a nonequilibrium steady state (NESS). Thus, biological systems are often better characterized as nonequilibrium steady states rather than equilibrium systems. Such NESS may change in response to changing environmental conditions.  Given that selective advantage may be incurred by energetically-efficient operation, evolution may have sculpted biological components to interact so as to reduce the energy wasted during transitions between NESS. Accordingly, optimizing such transitions may offer insights into the design principles of biological systems and guide the creation of synthetic molecular-scale machines.

Much recent attention has focused on predicting optimal protocols to drive systems between equilibrium states with minimal expended work~\cite{Sivak2012b,Zulkowski2012,Shenfeld,Brody2009,Seifert2008,Seifert2007,Aurell}. In particular, Ref.~\cite{Sivak2012b} proposes a linear response framework for protocols that minimize the dissipation during nonequilibrium perturbations of microscopic systems. This idea is developed further in~\cite{Zulkowski2012} where the utility of Riemannian geometry suggested in~\cite{Sivak2012b} is exploited to find explicit optimal protocols for a paradigmatic colloidal particle model. Our ultimate aim is to extend the geometric framework of~\cite{Sivak2012b} to optimal transitions between steady states so that the tools utilized in~\cite{Zulkowski2012} may be applied to more biologically relevant models.

In this article, we take a first step towards this goal by optimizing 
 the Hatano-Sasa Y-value, a quantity similar to dissipated work, for the paradigmatic model system tested in~\cite{Trepagnier2004} and analyzed in~\cite{Mazonka_1999} with an eye towards experimental tests. We calculate closed-form expressions for both the geodesic optimal protocol and the optimal straight-line protocol and test these protocols numerically via a system of equations derived from the Fokker-Planck equation. Finally, we propose a regime of validity of our approximation based on this numerical work. By measuring the average work required to drive this system along either optimal or naive paths through control parameter space, our results can be tested experimentally in a straightforward way using existing experimental techniques.

\section*{Results}\label{results}

\subsection*{The model system and its inverse diffusion tensor}\label{model_system}
We consider a particle with spatial coordinate $x$ diffusing under Langevin dynamics subject to a one-dimensional harmonic potential, with equation of motion
\begin{equation}
\label{eom}
\dot{x} = -\frac{k(t)}{\gamma} x  + \eta(t) - v(t) \ ,
\end{equation}
for Gaussian white noise $ \eta(t)$ satisfying
\begin{equation}
\left\langle \eta(t) \right\rangle = 0 \ , \quad \left< \eta(t) \eta(t') \right> = \frac{2 }{\beta \gamma} \delta (t-t') \ .
\end{equation}
Here $\gamma$ is the Cartesian friction coefficient and $ x $ is the coordinate of the colloidal particle in the frame co-moving with the trap.
The particle is initially in NESS due to constant trap velocity $ v $.

As defined in~\cite{Hatano2001}, the Hatano-Sasa Y-value
\begin{equation}\label{EEP}
Y \equiv \int_{0}^{\tau} dt \ \bigg[ \frac{ d \boldsymbol \lambda^{T} }{dt} \bigg] \cdot \frac{\partial \phi}{\partial \boldsymbol \lambda}\big(x(t);\boldsymbol \lambda(t)\big) 
\end{equation}
arises in NESS transitions when the control parameters $\boldsymbol \lambda$ are changed rapidly compared to the system's relaxation timescale. Here $ \phi(x;\boldsymbol \lambda) \equiv - \ln \rho_{ss}(x; \boldsymbol \lambda) $ where $\rho_{ss}(x; \boldsymbol \lambda) $ is the steady state probability distribution and $ \tau $ is the protocol duration. 

In some simple cases this corresponds to the system `lagging' behind the changing control parameters. For transitions between equilibrium states this measure reduces to the standard dissipation governed by the Clausius inequality~\cite{Kardar}. This measure of irreversibility~\eqref{EEP} obeys a significant NESS fluctuation theorem that has been experimentally observed in our particular model system~\cite{Trepagnier2004}. We may derive an approximate scheme, exact in the linear response regime~\cite{Zwanzig2001}, for optimizing this Y-value during finite-time transitions between different nonequilibrium states.

The ensemble average of the Y-value is
\begin{equation}
\label{Ydefn}
\langle Y \rangle_{\boldsymbol \Lambda} \equiv \int_{0}^{\tau} dt \ \bigg[ \frac{ d \boldsymbol \lambda^{T} }{dt} \bigg] \cdot \bigg< \frac{\partial \phi}{\partial \boldsymbol \lambda}\big(x;\boldsymbol \lambda(t)\big) \bigg>_{\boldsymbol \Lambda} \ .
\end{equation}
During the driving process, the system's probability distribution over microstates fundamentally depends on the history of the control parameters $\boldsymbol\lambda $, which we denote by the control parameter protocol $\boldsymbol \Lambda$. We assume the protocol to be sufficiently smooth to be twice-differentiable.

Applying linear response theory~\cite{Zwanzig2001,Sivak2012b,Zulkowski2012} and assuming that the protocol varies sufficiently slowly~\cite{Sivak2012b}, we arrive at an expression for the average Y-value
\begin{equation}
\label{Yapprox}
\langle Y \rangle_{\boldsymbol \Lambda} \approx \int_{0}^{\tau} dt \ \bigg[ \frac{ d \boldsymbol \lambda^{T} }{dt} \bigg] \cdot \boldsymbol \zeta(\boldsymbol \lambda(t)) \cdot \bigg[ \frac{ d \boldsymbol \lambda }{dt} \bigg] \ ,
\end{equation}
in terms of the control parameter velocities $d{\boldsymbol \lambda}/dt$ and the inverse diffusion matrix $\zeta({\boldsymbol \lambda})$ with entries
\begin{equation}\label{invdiffdefn} \zeta_{ij}(\boldsymbol \lambda) \equiv \int_{0}^{\infty} dt' \ \bigg< \frac{\partial \phi}{\partial \lambda^{i}}(t') \ \frac{\partial \phi}{\partial \lambda^{j}}(0) \bigg>_{\boldsymbol \lambda}. \end{equation}
The angle brackets $ \langle \dots \rangle_{\boldsymbol \lambda} $ represent an average over noise followed by a stationary state average over initial conditions using the distribution $ \rho_{ss}(x; \boldsymbol \lambda) $. Note that, if $ \rho_{ss}(x;\boldsymbol \lambda) $ is the equilibrium distribution, the inverse diffusion tensor of~\cite{Zulkowski2012} is recovered.

In general, the detailed balance condition is violated in NESS and so the matrix $ \big< \partial_{\lambda^{i}} \phi(t') \partial_{\lambda^{j}} \phi(0) \big>_{\boldsymbol \lambda} $ may be asymmetric. Eq.~\eqref{Yapprox} shows the use of Eq.~\eqref{invdiffdefn} (specifically its symmetric part) as a metric tensor is not precluded. However,$ \big< \partial_{\lambda^{i}} \phi(t') \partial_{\lambda^{j}} \phi(0) \big>_{\boldsymbol \lambda} $ is not a covariance matrix and so a general proof of positive-definiteness is lacking~\cite{Sivak2012b}. These considerations do not affect the model considered here but future work is needed to address this issue for the general case.

The steady-state distribution is given by $ \rho_{ss}(x; \boldsymbol \lambda) \equiv \sqrt{\frac{\beta k}{2 \pi}} \exp \{ - \frac{\beta}{2 k } \left( k x + \gamma v \right)^2 \} $~\cite{Trepagnier2004,Mazonka_1999}. The parameter space derivative of $ \phi $ is given by
\begin{subequations}
\begin{align}
\frac{\partial \phi}{\partial \boldsymbol \lambda}  & \equiv \bigg( \frac{\partial \phi}{\partial k}, \frac{\partial \phi}{\partial v} \bigg) \\
& = \bigg(-\frac{1}{2k}+\frac{\beta}{2} x^2-\frac{\beta}{2} \left[ \frac{ \gamma v}{k} \right]^2 ,  \beta \gamma \left[ x+ \frac{ \gamma v}{k} \right] \bigg) \ .
\end{align}
\end{subequations}
In order to calculate the time correlation functions in Eq.~\eqref{invdiffdefn}, we solve Eq.~\eqref{eom} for constant $ k $ and $ v $, giving
\begin{equation}
\label{xsoln}
x(t) = x_{0} e^{-\frac{k}{\gamma} t} + \int_{0}^{t} ds \ e^{-\frac{k}{\gamma} (t-s)} \eta(s) -\frac{\gamma v}{k} \left( 1-e^{-\frac{k}{\gamma} t} \right) \ .
\end{equation}
Recalling that $ \eta(t) $ is Gaussian noise, Eq.~\eqref{xsoln} implies
\begin{subequations}
\begin{align}
\big< \partial_{k} \phi(t) \partial_{k} \phi(0) \big>_{\boldsymbol \lambda} & = \frac{ \beta \left( \gamma v \right)^2}{k^3} e^{-\frac{k}{\gamma} t} + \frac{1}{2 k^2} e^{-\frac{2 k}{\gamma} t} \ , \\
\big< \partial_{k} \phi(t) \partial_{v} \phi(0) \big>_{\boldsymbol \lambda} & = -\beta v \left( \frac{\gamma}{k} \right)^2 e^{-\frac{k}{\gamma} t} \ , \\
\big< \partial_{v} \phi(t) \partial_{k} \phi(0) \big>_{\boldsymbol \lambda} & = -\beta v \left( \frac{\gamma}{k} \right)^2 e^{-\frac{k}{\gamma} t} \ , \\ 
\big< \partial_{v} \phi(t) \partial_{v} \phi(0) \big>_{\boldsymbol \lambda} & = \frac{\beta \gamma^2}{k} e^{-\frac{k}{\gamma} t} \ .
\end{align}
\end{subequations}
Integrating over time yields the inverse diffusion tensor:
\begin{equation}
\label{tensor}
\boldsymbol \zeta(k,v) =
\left(
\begin{array}{cc}
\frac{\gamma}{4 k^4} \left[ k + 4 \beta \left( \gamma v \right)^2 \right] & -\beta v \left[ \frac{\gamma}{k} \right]^3 \\
-\beta v \left[ \frac{\gamma}{k} \right]^3 & \beta \frac{\gamma^3}{k^2} \\
\end{array}
\right) \ .
\end{equation}

\subsection*{Optimal protocols}
Though one can write down the geodesic equations for the metric [Eq.~\eqref{tensor}] in the $(k,v)$ coordinate system, more insight is gained by finding a suitable change of coordinates. A direct calculation of this metric's Ricci scalar yields $ R = 0$, demonstrating that the underlying geometry is Euclidean~\cite{doCarmo}.

The line element corresponding to the metric in Eq.~\eqref{tensor} is
\begin{align}
\label{lineelem}
ds^2 = \frac{\gamma}{4 k^4} &\left[ k + 4 \beta \left( \gamma v \right)^2 \right]  dk^2 \\
&- 2 \beta v \left( \frac{\gamma}{k} \right)^3 dk \ dv + \beta \frac{\gamma^3}{k^2} dv^2 \ . \notag
\end{align}
To find the explicit coordinate transformation making the Euclidean geometry manifest, we write the line element as
\begin{equation}
ds^2 = \beta \gamma^3 \bigg\{ \bigg[ d \left( \frac{v}{k} \right) \bigg]^2 + \bigg( \frac{d k}{2 \sqrt{\beta} \gamma k^{\frac{3}{2}}} \bigg)^2 \bigg\} \ .
\end{equation}
This suggests the coordinate transformation $ \xi = \frac{v}{k} \ , \ \chi = \frac{1}{\gamma \sqrt{ \beta k}} $, so that
\begin{equation}
\label{Euclidelem}
ds^2 = \beta \gamma^3 \big( d\xi^2 + d\chi^2 \big) \ .
\end{equation}

In this coordinate system, geodesics are straight lines of constant speed. To find optimal protocols in $ (k,v) $ space, one simply transforms the coordinates of the endpoints into $ (\xi,\chi) $ space, connects these points by a straight line, and uses the inverse transformation to map the line onto a curve in $ (k,v)$ space. This follows from the invariance of the geodesic equation~\cite{doCarmo}. Explicitly, the optimal protocol joining $(k_{i},v_{i})$ and $ (k_{f},v_{f}) $ is
\begin{subequations}\label{geo}
\begin{align}
k(t) & = \bigg[ \frac{1}{\sqrt{k_{i}}}(1-T)+\frac{1}{\sqrt{k_{f}}} T \bigg]^{-2}, \\
v(t) & = k(t) \bigg[\frac{v_{i}}{k_{i}}(1-T) + T \frac{v_{f}}{k_{f}} \bigg] \ ,
\end{align}
\end{subequations}
where $ T = \frac{t}{\tau} $. Sample optimal protocols are pictured in Fig.~\ref{fig:geodesics}.

\subsection*{Optimal straight-line protocols}
In the absence of any particular information about the system's dynamical properties, a naive control strategy would change the control parameters at a constant rate, producing a straight line in control parameter space. The inverse diffusion tensor approximation [Eq.~\eqref{Yapprox}] provides a recipe for choosing both  a potentially nonlinear path through control parameter space, as well as a time-course along that path. The inverse diffusion tensor formalism can alternatively be used to optimize the time-course along a straight-line control parameter path. Such a protocol provides a benchmark against which an optimal protocol [Eq.~\eqref{geo}] can be compared. For the model considered here, we will find that an optimal straight-line protocol can be substantially better than the most naive (constant-speed) straight-line protocol. Furthermore, straight-line protocols are relatively straightforward to test experimentally.

\begin{figure}[!ht]
\begin{center}
\includegraphics{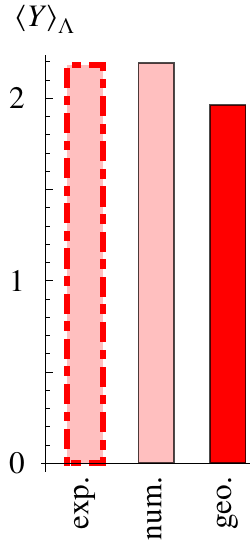} \
\includegraphics{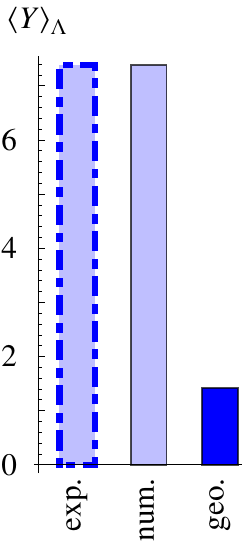} \
\includegraphics{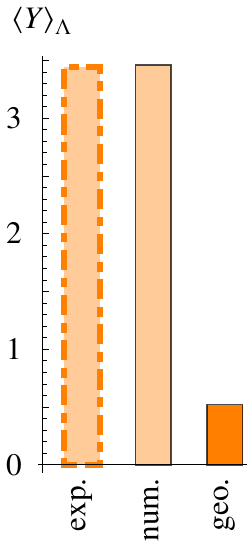}
\\ 
\hspace{1em} Exp. 1  \hspace{5em} Exp. 2 \hspace{5 em} Exp. 3 \\
\end{center}
\caption{{\bf Optimal protocols outperform constant-$k$ protocols tested in Ref.~\cite{Trepagnier2004}.} Experiment 1 (left, red) used a quarter-sine wave protocol to vary the trap speed; Experiments 2 (middle, blue) and 3 (right, orange) used an inverted three-quarters sine wave. Specifically, $ v(t) = 8.12 + 4.03 \sin{ \left( \pi t / 2 \tau \right) } ,k =4.25, \tau = 0.06, q = 0.20$ for Experiment 1, $ v(t) = 9.93 - 3.63 \sin{  \left( 3 \pi t / 2 \tau \right) } ,k =4.51, \tau = 0.06, q = 0.21 $ for Experiment 2, and $ v(t) = 7.53 - 2.67 \sin{  \left( 3 \pi t / 2 \tau \right) } ,k =4.9, \tau = 0.08, q=0.23  $ for Experiment 3. Here, velocity is measured in $ \mu\mbox{m}/s $, $ \tau $ is the protocol duration measured in $s$, $ k $ is the trap stiffness measured in $\mbox{pN} / \mu\mbox{m} $, and $ q \equiv k / \beta \gamma$ is measured in $\mbox{pN} \ \mu\mbox{m} /s $ . The Y-value for these protocols (light color bar) and for the optimal protocols (solid color bar) were obtained numerically assuming $ \beta^{-1} = 4.6~\mbox{pN}~\mbox{nm}$ (red), $ \beta^{-1} = 4.45~\mbox{pN}~\mbox{nm}$ (blue), $ \beta^{-1} = 4.35~\mbox{pN}~\mbox{nm}$ (orange) respectively.
These effective temperatures were chosen to give the best match between experiment and numerical calculation, and
may differ from room temperature ($ \beta^{-1} = 4.14~\mbox{pN}~\mbox{nm}$) because of local heating by the optical trap~\cite{Peterman2003}. We predict a significant reduction in Y-value for optimal protocol driving under the conditions of the three experiments described in Ref.~\cite{Trepagnier2004}.}
\label{fig:exp}
\end{figure}

When $ k(t) $ is held fixed, a straightforward application of variational calculus demonstrates that a straight-line protocol in $ v(t) $ is \emph{exactly} optimal and agrees with the predictions of the linear response approximation [Eq.~\eqref{Yapprox}]. In Ref.~\cite{Trepagnier2004}, the average Y-value was measured for three distinct experimental trials involving protocols with constant $ k $. As summarized in Fig.~\ref{fig:exp}, the optimal protocol, namely the naive straight line in the case of constant $ k $, shows significantly reduced Y-value compared with the protocols used in each experimental trial. However, in terms of testing the performance of the optimal protocols [Eq.~\eqref{geo}], $ k_{f} \neq k_{i} $ is the more general case.

As in the case of finding globally optimal protocols, the problem of finding optimal straight line protocols simplifies dramatically in $ (\xi,\chi) $ coordinates. Using Eq.~\eqref{Euclidelem}, we find
\begin{equation}
\big< Y \big>_{\boldsymbol \Lambda} \approx \beta \gamma^3 \int_{0}^{\tau} dt \bigg[ 1+ b^2 \chi^2(t) \bigg] \left( \frac{d \chi}{d t} \right)^2 \ ,
\end{equation}
for
\begin{equation}
b \equiv 2\beta \gamma^2\frac{k_f v_i - k_i v_f}{k_f-k_i} \ .
\end{equation}
The Euler-Lagrange equation implies
\begin{equation}
\frac{d \chi}{dt} = \frac{ \frac{1}{\tau} \int_{\chi_{0}}^{\chi_{f}} dz \ \sqrt{1+ b^2 z^2 }  }{\sqrt{ 1+ b^2 \chi^2(t) }} \ ,
\end{equation}
which determines an implicit expression for $ \chi(t) $:
\begin{align}
& 2b \left( \frac{t}{\tau} \right) \int_{\chi_{0}}^{\chi_{f}} dz \ \sqrt{1+ b^2 z^2 } =  b \bigg( \chi(t) \sqrt{1+ b^2 \chi^2(t) } \nonumber\\ &-\chi_{0} \sqrt{1+ b^2 \chi_{0}^2} \bigg) + \sinh^{-1} \left[ b \chi(t) \right] - \sinh^{-1} \left[ b \chi_{0} \right] \ .
\end{align}
The relation $ \chi = \frac{1}{\gamma \sqrt{ \beta k }} $ determines an implicit expression for $ k(t) $, and hence for $ v(t) $.

\subsection*{Computing the Y-value numerically}
We validate the optimality of the geodesics [Eq.~\eqref{geo}] and compare with optimal straight-line protocols by calculating the average Y-value directly by integrating in time the Fokker-Planck equation describing the dynamical evolution of the particle probability distribution~\cite{Mazonka_1999},
\begin{equation}
\label{FokkerPlanck}
\frac{\partial \rho}{\partial t} = \frac{k(t)}{\gamma} \frac{\partial }{\partial x} \left( x \rho \right) + v(t) \frac{\partial \rho}{\partial x} + \frac{1}{\beta \gamma} \frac{\partial^2 \rho}{\partial x^2} \ . \end{equation}
In full generality, the mean Y-value as a functional of the protocol $ \boldsymbol \lambda (t) = (k(t),v(t)) $ is
\begin{align}
\label{expY}
\int_{0}^{\tau} dt \Bigg[ - \frac{ \dot{k}}{2 k } &- \frac{\beta}{2} \left( \frac{\gamma v}{k} \right)^2 \dot{k}+ \frac{\beta}{2} \dot{k} \langle x^2 \rangle_{\boldsymbol \Lambda} \notag\\
&+ \beta \gamma \dot{v} \langle x \rangle_{\boldsymbol \Lambda} + \beta \gamma^2 \frac{v}{k} \dot{v} \Bigg] \ .
\end{align}
Here angled brackets denote averages over the nonequilibrium probability density $ \rho(x,t) $.

By integrating Eq.~\eqref{FokkerPlanck} against $ x $  and $ x^2 $, we find a system of equations for relevant nonequilibrium averages:
 \begin{subequations}
 \label{FPsys}
\begin{align}
 \frac{d}{dt} \langle x \rangle_{\boldsymbol \Lambda}   &=  -\frac{k(t)}{\gamma}\langle x \rangle_{\boldsymbol \Lambda} - v(t) \ ,  \\
 \frac{d}{dt} \langle x^2 \rangle_{\boldsymbol \Lambda}   &= -\frac{2 k(t)}{\gamma} \langle x^2 \rangle_{\boldsymbol \Lambda} - 2 v(t) \langle x \rangle_{\boldsymbol \Lambda} + \frac{2}{\beta \gamma}\ , \
 \end{align}
\end{subequations}
supplemented by initial conditions
\begin{subequations}
\label{FPsysinitconds}
\begin{align}
\left< x \right>_{\boldsymbol \Lambda}(0) &= -\frac{\gamma v_{i}}{k_{i}} \ , \\
\left< x^2 \right>_{\boldsymbol \Lambda}(0) &= \frac{1}{\beta k_{i}}+ \left[ \frac{\gamma v_{i}}{k_{i}} \right]^2 \ . \
\end{align}
\end{subequations}
Note that for a more complex system the first and second moments $\langle x\rangle$ and $\langle x^2\rangle$ are not sufficient to characterize the probability distribution, but time-dependent solutions are still accessible through standard (but more computationally intensive) numerical integration of the full Fokker-Planck equation~\eqref{FokkerPlanck}~\cite{Risken1996}.

\begin{figure}[t]
\begin{center}
\includegraphics{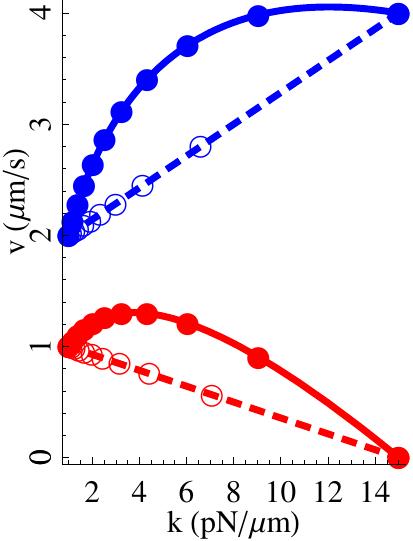} \
\includegraphics{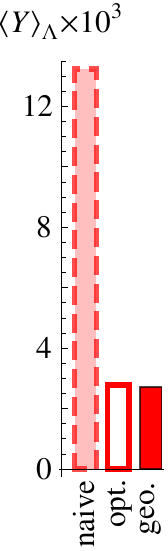} \
\includegraphics{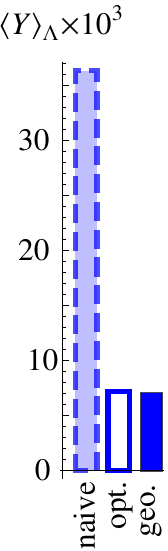}
\\ (a) \hspace{10em} (b) \\
\end{center}
\caption{{\bf Geodesics describe protocols that outperform naive (constant-speed) straight-line paths in parameter space.} Geodesics between fixed pairs of points in the $ (k,v)$-plane and accompanying straight-line protocols are pictured in (a). The filled circles represent points separated by equal times. The open circles correspond to the optimal parametrization along the respective straight path. All mean Y-values were calculated using the Fokker-Planck system, Eq.~\eqref{FPsys}.
Here, $ \gamma = 0.1~\mbox{pN}~\mbox{s} / \mu\mbox{m} $ and $ \beta^{-1} = 4.6~\mbox{pN}~\mbox{nm}$ to approximate the experiments of Ref.~\cite{Trepagnier2004}. The protocol duration is chosen to be $ \tau = 100~s$
to ensure that the relative error $\big| 1-\langle Y \rangle_{\boldsymbol \Lambda}^{\rm approx}/ \langle Y \rangle_{\boldsymbol \Lambda} \big| $ is less than $ 1.4\% $ for all protocols. Protocol endpoints were selected for experimental accessibility~\cite{Block2004}. The relative performance of naive straight-line, optimal straight-line, and geodesic protocols are summarized in (b).}
\label{fig:geodesics}
\end{figure}

We solve these equations numerically and compare the performance of optimal straight lines against geodesics [Eq.~\eqref{geo}] and naive (constant-speed) straight-line protocols in Fig.~\ref{fig:geodesics}. We selected endpoints and physical constants based on those used in the experiments of Ref.~\cite{Trepagnier2004} which may be found in the caption of Fig.~\ref{fig:exp}. In this near-equilibrium regime the inverse diffusion tensor approximation produces small relative error in Y-value. Though there is only a marginal difference in performance between the optimal straight-line protocol and the geodesic for both sets of endpoints, there is a substantial benefit in using either over the naive straight line protocol.

\subsection*{The inverse diffusion tensor arises naturally from the Fokker-Planck equation.}

If we neglect terms involving second- and higher-order temporal derivatives (an alternative near-equilibrium approximation), we obtain an approximate solution to the Fokker-Planck system:
\begin{subequations}
\begin{align}
\left< x \right>_{\boldsymbol \Lambda} &\approx -\gamma \frac{v}{k} + \left( \frac{\gamma}{k} \right)^2 \dot{v} - \frac{v}{k} \left( \frac{\gamma}{k} \right)^2\dot{k} \ ,  \\
\left< x^2 \right>_{\boldsymbol \Lambda} &\approx \gamma^2 \frac{v^2}{k^2} + \frac{1}{\beta k} + \frac{\gamma \dot{k}}{2 \beta k^3}+ \frac{2 \gamma^3 v^2 \dot{k}}{k^4} -\frac{2 \gamma^3 v \dot{v}}{k^3} \ . \
\end{align}
\end{subequations} \\
Substituting this into the expression for the mean Y-value [Eq.~\eqref{expY}], we recover Eq.~\eqref{lineelem}. The argument above suggests that the emergence of the inverse diffusion tensor from the Fokker-Planck equation may follow from a perturbation expansion in small parameters \cite{Zulkowski2012}.\\

\section*{Discussion}
We have taken the first step towards extending the geometric framework for calculating optimal protocols presented in~\cite{Sivak2012b,Zulkowski2012} to systems relaxing to NESS. As energy-transducing biological systems are more faithfully described by NESS than by equilibrium statistics, this brings recent theoretical developments closer to the behavior of \emph{in vivo} biological systems.

Using a linear response approximation, we found the optimal mean Y-value for a model system of a colloidal particle (initially in NESS) dragged through solution and subject to a time-dependent harmonic potential. We took as our control parameters the velocity and spring constant of the harmonic potential. As in~\cite{Zulkowski2012}, tools from Riemannian geometry revealed a useful coordinate transformation which greatly simplified the construction of optimal straight-line protocols as well as geodesic protocols. These optimal protocols were tested numerically and the small relative error in the Y-value approximation for experimentally relevant choices of parameters is encouraging.

Our predictions may be tested experimentally with existing hardware and methods. In Ref.~\cite{Trepagnier2004}, the authors report on experiments performed with micron-sized polystyrene beads in solution. The harmonic potential is created by superposing the foci of two counterpropagating laser beams. The location of this trap was translated using a steerable mirror. The velocity $v$ of the trap location was altered by changing the mirror's angular rate of rotation, and the trap stiffness $k$ can be manipulated by dynamically changing the intensity of the laser beam~\cite{Joykutty2005} or by passing the laser beam through a polarization filter and dynamically changing the polarization of the laser beam. Force is inferred from the rate of change of the momentum of light measured by position-sensitive photodetectors. Comparison of the average work incurred during different protocols would provide an experimental test of the optimal protocols predicted in this manuscript.

Using the inverse diffusion tensor approximation in general allows us access to the full power of Riemannian geometry in calculating optimal protocols. However, such experimental tests can assess the range of validity of the approximation. Our alternate derivation of the inverse diffusion tensor via a ``derivative-truncation" expansion~\cite{Zulkowski2012} suggests a greater robustness of the approximation.

In this paper we provide concrete theoretical predictions for experiments --- specifically, we find that geodesics, optimal straight-lines, and naive straight-line protocols all are substantially more efficient than the protocols tested in Trepagnier, \emph{et al.} Moreover, we demonstrate that for simultaneous adjustment of $k$ and $v$, optimal straight-line protocols can perform substantially better than naive (constant-speed) straight-line protocols. The necessary methodology and experimental apparatus are well-established~\cite{Trepagnier2004} to not only verify our predictions but to push beyond the near-steady-state regime.

Given the greater generality embodied by the extension to NESS, and the accuracy of this approximation for a standard model system, optimal driving protocols derived in this framework promise greater applicability to models of biomolecular machines. 
Nevertheless, important hurdles remain: our model system experiences forces linear in position and has a steady-state distribution differing from the equilibrium one only in its average displacement. Molecular machines feature nonlinear force profiles, potentially nontrivial steady-state distributions, and often operate far from equilibrium. Thus our comparatively simple theoretical framework may need further elaboration to address the dynamics and efficiency of molecular machines with reasonable fidelity. 

Furthermore, the relatively simple model system we treat in this manuscript represents a new frontier for the analytical solution of optimal protocols under the inverse diffusion tensor approximation. For significantly more complicated models of greater biological interest, a simple general approach (in lieu of a search for an analytical solution) would be a fully numerical method, involving the calculation of the inverse diffusion tensor at a grid of points in control parameter space, analogous to the approach in \cite{Shenfeld2009}.

Finally, there remains the important open question of what quantity or quantities are to be optimized in faithful models of biological processes. In this paper, we made the choice of optimizing the Y-value which has been experimentally studied in this particular model system~\cite{Trepagnier2004} and may be optimized by the same geometric framework as in~\cite{Sivak2012b}. These qualities were advantageous to begin a clear and mathematically tractable first step towards optimization of steady state transitions.

However, it is possible and perhaps likely that a properly defined average dissipated heat will be the biologically relevant quantity to optimize rather than the Y-value.  We anticipate that a geometric approach to optimization will be applicable to these more general systems and notions of heat production in a relevant regime of parameter values and protocol durations. However, a more general construction will have to take into account the so-called housekeeping heat~\cite{Hatano2001,SeifertReview_2012} which is generated in maintaining the steady state at given control parameter values. Future work is needed to address these issues properly. 


\section*{Acknowledgments}
We thank Christopher Jarzynski for a useful discussion of this work, and Vladislav Belyy, Ahmet Yildiz, and Jeff Moffitt for information regarding optical traps. M.\ R.\ D.\ gratefully acknowledges support from the McKnight Foundation, the Hellman Family Faculty Fund, the McDonnell Foundation, and the Mary Elizabeth Rennie Endowment for Epilepsy Research. M.\ R.\ D.\ and P.\ R.\ Z.\ were partly supported by the National Science Foundation through Grant No. IIS-1219199. D.\ A.\ S.\ was supported by NIGMS Systems Biology Center grant P50 GM081879.


\bibliography{Ycontrol}

\begin{thebibliography}{10}
\providecommand{\url}[1]{\texttt{#1}}
\providecommand{\urlprefix}{URL }
\expandafter\ifx\csname urlstyle\endcsname\relax
  \providecommand{\doi}[1]{doi:\discretionary{}{}{}#1}\else
  \providecommand{\doi}{doi:\discretionary{}{}{}\begingroup
  \urlstyle{rm}\Url}\fi
\providecommand{\bibAnnoteFile}[1]{%
  \IfFileExists{#1}{\begin{quotation}\noindent\textsc{Key:} #1\\
  \textsc{Annotation:}\ \input{#1}\end{quotation}}{}}
\providecommand{\bibAnnote}[2]{%
  \begin{quotation}\noindent\textsc{Key:} #1\\
  \textsc{Annotation:}\ #2\end{quotation}}
\providecommand{\eprint}[2][]{\url{#2}}

\bibitem{Schrodinger:WhatIsLife}
Schr\"odinger E (1992) What is Life?
\newblock Cambridge: Cambridge University Press.
\bibAnnoteFile{Schrodinger:WhatIsLife}

\bibitem{Howard}
Howard J (2001) Mechanics of Motor Proteins and the Cytoskeleton.
\newblock Sunderland, Massachusetts: Sinauer.
\bibAnnoteFile{Howard}

\bibitem{MBOC}
Alberts B, Johnson A, Lewis J, Raff M, Roberts K, et~al. (2002) Molecular
  Biology of the Cell.
\newblock New York: Garland Science.
\bibAnnoteFile{MBOC}

\bibitem{Sivak2012b}
Sivak DA, Crooks GE (2012) Thermodynamic metrics and optimal paths.
\newblock Phys Rev Lett 108: 190602.
\bibAnnoteFile{Sivak2012b}

\bibitem{Zulkowski2012}
Zulkowski PR, Sivak DA, Crooks GE, DeWeese MR (2012) Geometry of thermodynamic
  control.
\newblock Phys Rev E 86: 041148.
\bibAnnoteFile{Zulkowski2012}

\bibitem{Shenfeld}
Shenfeld DK, Xu H, Eastwood MP, Dror RO, Shaw DE (2009) Minimizing
  thermodynamic length to select intermediate states for free-energy
  calculations and replica-exchange simulations.
\newblock Phys Rev E 80: 046705.
\bibAnnoteFile{Shenfeld}

\bibitem{Brody2009}
Brody DC, Hook DW (2009) Information geometry in vapour-liquid equilibrium.
\newblock J Phys A 42: 023001.
\bibAnnoteFile{Brody2009}

\bibitem{Seifert2008}
Gomez-Marin A, Schmiedl T, Seifert U (2008) Optimal protocols for minimal work
  processes in underdamped stochastic thermodynamics.
\newblock J Chem Phys 129: 024114.
\bibAnnoteFile{Seifert2008}

\bibitem{Seifert2007}
Schmiedl T, Seifert U (2007) Optimal finite-time processes in stochastic
  thermodynamics.
\newblock Phys Rev Lett 98: 108301.
\bibAnnoteFile{Seifert2007}

\bibitem{Aurell}
Aurell E, Mej\'{i}a-Monasterio C, Muratore-Ginanneschi P (2011) Optimal
  protocols and optimal transport in stochastic thermodynamics.
\newblock Phys Rev Lett 106: 250601.
\bibAnnoteFile{Aurell}

\bibitem{Trepagnier2004}
Trepagnier EH, Jarzynski C, Ritort F, Crooks GE, Bustamante CJ, et~al. (2004)
  Experimental test of hatano and sasa's nonequilibrium steady-state equality.
\newblock Proc Natl Acad Sci USA 101: 15038-15041.
\bibAnnoteFile{Trepagnier2004}

\bibitem{Mazonka_1999}
Mazonka O, Jarzynski C (1999) Exactly solvable model illustrating
  far-from-equilibrium predictions.
\newblock ArXiv:9912.121.
\bibAnnoteFile{Mazonka_1999}

\bibitem{Hatano2001}
Hatano T, Sasa S (2001) {Steady-state thermodynamics of Langevin systems}.
\newblock Phys Rev Lett 86: 3463--3466.
\bibAnnoteFile{Hatano2001}

\bibitem{Kardar}
Kardar M (2007) Statistical physics of particles.
\newblock Cambridge: Cambridge University Press.
\newblock \urlprefix\url{http://dx.doi.org/10.1017/CBO9780511815898}.
\bibAnnoteFile{Kardar}

\bibitem{Zwanzig2001}
Zwanzig R (2001) Nonequilibrium statistical mechanics.
\newblock New York: Oxford University Press.
\bibAnnoteFile{Zwanzig2001}

\bibitem{doCarmo}
do~Carmo MP (1992) Riemannian geometry.
\newblock Boston: Birkh\"{a}user.
\bibAnnoteFile{doCarmo}

\bibitem{Risken1996}
Risken H (1996) The Fokker-Planck equation.
\newblock Berlin: Springer-Verlag, 3rd edition.
\bibAnnoteFile{Risken1996}

\bibitem{Joykutty2005}
Joykutty J, Mathur V, Venkataraman V, Natarajan V (2005) {Direct Measurement of
  the Oscillation Frequency in an Optical-Tweezers Trap by Parametric
  Excitation}.
\newblock Phys Rev Lett 95: 193902.
\bibAnnoteFile{Joykutty2005}

\bibitem{Shenfeld2009}
Shenfeld DK, Xu H, Eastwood MP, Dror RO, Shaw DE (2009) {Minimizing
  thermodynamic length to select intermediate states for free-energy
  calculations and replica-exchange simulations}.
\newblock Physical Review E 80: 046705.
\bibAnnoteFile{Shenfeld2009}

\bibitem{SeifertReview_2012}
Seifert U (2012) Stochastic thermodynamics, fluctuation theorems and molecular
  machines.
\newblock Reports on Progress in Physics 75: 126001.
\bibAnnoteFile{SeifertReview_2012}

\bibitem{Peterman2003}
Peterman EJ, Gittes F, Schmidt CF (2003) Laser-induced heating in optical
  traps.
\newblock Biophysical Journal 84: 1308 - 1316.
\bibAnnoteFile{Peterman2003}

\bibitem{Block2004}
Neuman K, Block S (2004) Optical trapping.
\newblock Review of Scientific Instruments 75: 2787-2809.
\bibAnnoteFile{Block2004}

\end{thebibliography}




\end{document}